\documentclass[twocolumn,superscriptaddress,showpacs,prl,amsmath,amssymb]{revtex4}
\usepackage{graphicx}
\usepackage{bm}
\begin{document}

\title{Generalized survival in equilibrium step fluctuations}

\author{ M. Constantin}
\affiliation{
Condensed Matter Theory Center,
Department of Physics, University of Maryland, College Park, Maryland 20742-4111
}
\affiliation{Materials Research Science and Engineering Center,
Department of Physics, University of Maryland, College Park, MD 20742-4111
}
\author{ S. \surname{Das Sarma}}
\affiliation{
Condensed Matter Theory Center,
Department of Physics, University of Maryland, College Park, Maryland 20742-4111
}

\begin{abstract}
We investigate the dynamics of a generalized survival probability 
$S(t,R)$ defined with respect to an arbitrary reference
level $R$ (rather than the average) in equilibrium step 
fluctuations. The exponential decay at large time scales of 
the generalized survival probability is numerically analyzed. 
$S(t,R)$ is shown to exhibit simple scaling behavior as a 
function of system-size $L$, sampling time $\delta t$, and 
the reference level $R$. The generalized survival time scale, 
$\tau_s(R)$, associated with $S(t,R)$ is shown to decay 
exponentially as a function of $R$.
\end{abstract}


\maketitle

Controlling the stability of nanostructures is an important
fundamental issue in nanoscience. A key problem in this respect is
the random stochastic interface dynamics associated with equilibrium
nanometer scale thermal fluctuations unavoidably present in all
nanosystems. Very interesting questions 
\cite{krug,us1,magda1,expt2,BR} have recently been addressed 
regarding the complex dynamics of fluctuating interfaces in the 
context of first--passage statistics, which seems 
to be the appropriate framework in investigating the time 
it takes for a nanodevice to first fluctuate into an unstable 
state. It turns out that 
useful dynamical quantities such as persistence probability $P(t)$ 
\cite{krug,us1,expt2} (i.e. the probability that a stochastic variable
does not return to its initial value over time $t$) and survival
probability $S(t)$ \cite{us1,BR} (i.e. the probability that a
stochastic variable does not cross its average level up to time $t$) 
can be numerically and experimentally investigated for interfaces 
with dynamics governed by various kinetic mechanisms (such as
high--temperature attachment/detachment of atoms at the step edge 
or low--temperature step edge diffusion of atoms) to gain 
insight into the stability issue.  

Much work has been devoted \cite{rev2,ellen} over the last decade in 
understanding equilibrium fluctuations on vicinal surfaces, 
mostly using the dynamic scaling approach. If $h(x,t)$ is the 
dynamical height (with respect to the chosen reference 
position which is defined to be the average position, i.e. the 
$h=0$ line) fluctuation of a thermally fluctuating step as 
a function of the lateral position $x$ and time $t$ (where $t$ 
is also measured from an arbitrary time origin), then $h(t)$ 
at each value of $x$ is a stochastic dynamical variable by 
virtue of equilibrium thermal fluctuations. Because of the 
spatially extended nature of the step fluctuations through its 
dependence on both $x$ and $t$, the problem is non-Markovian, 
and persistence \cite{maj} and survival \cite{BR} concepts 
should be particularly relevant statistical tools in 
understanding the complex problem of surface fluctuations.

In this paper we introduce the new concept of a generalized survival
probability which enables us to probe deeper into the nature of the
stochastic process of interface step fluctuations. The 
generalized survival probability is the probability 
$S(t,R)$ that a given lateral step position $x$ with a height 
(i.e., step fluctuation measured from the equilibrium step position) 
$h(x,t)$ at time $t$ does never cross a pre-assisgned reference level
of the height, $R$, throughout the entire evolution. The particular
case with $R=0$ (i.e., the probability of the dynamical step height 
not returning in time $t$ to its average (``equilibrium'') $R=0$
level) has been studied recently \cite{BR} both analytically
and experimentally, and it has been shown to exhibit an exponential 
decay at large times, $S(t)\propto \exp(-t/\tau_s)$, where $\tau_s$ is the
survival time scale that provides information about the underlying
kinetics. The resulting surface step fluctuation survival probability 
$S(t)=S(t,R=0)$ and the associated time scale $\tau_s$ have also 
recently been studied experimentally using dynamical scanning
tunneling microscopy (STM) on different metallic systems: Al steps 
on Si $(111)$ surface at high temperatures, and Ag and Pb $(111)$ 
surfaces at relatively low temperatures \cite{future_exp}. 
In this paper we show numerically that $S(t,R)$ also 
has an exponential behavior at large time, 
$S(t,R)\propto \exp(-t/\tau_s(R))$, where $\tau_s(R)$ is the
generalized survival time scale. Our study reveals the dependence
of $\tau_s(R)$ on the system size $L$, sampling interval $\delta t$,
and reference level position $R$, allowing us to establish the complete
scaling form of $S(t,R)$. In particular, the sampling interval
(i.e. the time between successive measurements) \cite{samp} turns
out to be an essential ingredient inherent in any real experimental
measurement procedure. Also the study of the dependence of the
generalized survival time scale on the choice of the reference level 
$R$, which turns out to be exponential, should have 
particular importance for understanding the effect of thermal 
fluctuations at the nanoscale.

In this study we consider the case of the high-temperature step 
fluctuations dominated by atomistic attachment and detachment 
(AD), where the step edge is known \cite{rev2} to be well described 
by the coarse-grained second-order non-conserved linear Langevin 
equation, also known as the Edwards--Wilkinson (EW) equation \cite{EW}  
\begin{equation}
\label{EW}
\frac {\partial h(x, t)} {\partial t} = \nabla^{2} h(x,t) + \eta(x, t),
\end{equation}
\noindent where $\nabla^{2}$ refers to the spatial derivative 
(with respect to $x$, the lateral position along the step),
$\eta(x,t)$ with $\langle \eta(x, t) \eta(x^{'},t^{'}) \rangle 
= 2D \delta(x-x^{'})\delta(t-t^{'})$ is the usual uncorrelated 
random gaussian noise corresponding to the non-conserved white 
noise associated with the random AD process, and $D$ is the noise 
strength. The AD process, thought to be extremely important for
relatively high-temperature step fluctuations, has been extensively
studied in the literature using the EW equation \cite{ellen}. 

For equilibrium step fluctuations, we define the generalized survival
probability with respect to the height reference level $R$, $S(t,R)$, 
as the probability for the height variable to remain consistently 
$above$ a certain pre-assigned value ``$R$'' over time $t$:
\begin{equation}
\label{def}
S(t,R) \equiv \hbox{Prob}~\lbrace ~h(x,t^{\prime}) >~R,
~\forall ~ t_0 \leq t^{\prime} \leq t_0+t~ \rbrace,
\end{equation}
\noindent where $h(x,t)$ is the dynamical height of the interface 
at a fixed lateral position $x$ at time $t$, and $t_0$ is the 
initial time of the measurement. Although the above definition 
involves the dynamical variable $h(x,t)$ defined for a particular 
lateral position $x$, we take a statistical ensemble average 
over all lateral positions to obtain a purely time dependent 
stochastic dynamical quantity $S(t,R)$. Obviously, another quantity
that can be measured is the probability for the height stochastic
variable to remain $below$ the reference level up to time $t$. Since
in our case the dynamics of the interface fluctuations obeys a linear
stochastic equation, the interface preserves the up-down symmetry
along the direction perpendicular to the step edge. As a consequence,
in what follows we consider the average of the probabilities of
remaining always above $R$ and below $-R$, with $R \geq 0$.

The generalized survival probability function, $S(t,R)$, defined in 
Eq.~(\ref{def}) above, leads to a hierarchy of generalized survival 
time scales, $\tau_s(R)$, {\it if} the steady-state decay 
of $S(t,R)$ in time follows an exponential trend, 
$S(t,R) \sim e^{-t/\tau_{s}(R)}$. As we show below, this indeed
is obtained for Edwards--Wilkinson equilibrium step fluctuation 
phenomena, allowing us to define and measure the non-trivial survival 
time scale $\tau_s(R),\,0 \le R \le R_{max}$, that varies 
between $\tau_{s}(R=0)$ and $\tau_s(R_{max})$, where $\tau_s(R=0)$ 
is the usual survival time scale and $\tau_s(R_{max})$ is the survival
time with respect to the highest reference level $R_{max}$ that can be
defined for a model with finite roughness (i.e. rms fluctuations 
of the height variable with respect to the average). $R_{max}$ is
limited by the maximum value of the height fluctuation amplitude. 
Obviously, $S(t,R)$ and $\tau_s(R)$ are natural generalizations of 
the survival probability $S(t)$ and the survival time scale $\tau_s$, 
respectively, to the more complex concept of distribution of generalized 
survival times with limiting behavior (i.e. $R=0$) providing the 
usual survival time.

The exponential decay at large time of $S(t,R)$ that we find
numerically is not surprinsing. The generalized survival probability
with respect to the reference level $R$ can be regarded as the
probability $Z(t)$ of no zero crossing of the new stochastic variable
$H(x,t)=h(x,t)-R$. What we are looking for is the probability for the
stochastic variable $H(x,t)$ to remain positive up to time $t$ (or,
equivalently, the probability for $h(x,t)+R$ to remain negative over
time $t$). This type of question for the gaussian stationary processes
with zero mean has been addressed by mathematicians for a long time
\cite{Slepian}. The no zero crossing probability is traditionally
investigated in conjunction with the autocorrelation function,
$C_H\equiv \langle H(x,t_1)H(x,t_2)\rangle$ (where 
$\langle...\rangle$ represents an average over all realizations of
$H(x,t)$ arising from the thermal noise source). It is known 
\cite{NR} that for a stationary gaussian process (i.e. 
$C_H=f(|t_1-t_2|)\equiv f(t)$) with an autocorrelation 
function decaying faster than $1/t$ at large $t$, the asymptotic 
behavior of the no zero crossing probability is exponential, 
$Z(t)\propto \exp(-\mu t)$. The autocorrelation function
$C_H(t)$ itself has been shown \cite{BR} to be stationary at 
late times and to decay exponentially. This, along with the 
exponential decay of $Z(t)$, ensures an exponential decay for $S(t,R)$.  

In order to numerically simulate the process described by
Eq.~(\ref{EW}), we have used discrete stochastic Monte Carlo
simulations of the corresponding atomistic solid--on--solid model, the
extensively studied Family model \cite{rev2}, which belongs
asymptotically to the Edwards--Wilkinson universality class
\cite{EW}. The Family model in (1+1)--dimensions (i.e. one spatial
variable and one temporal variable) is characterized by $\beta=1/4$,
$\alpha=1/2$ and $z=\alpha/\beta=2$ \cite{rev2}, where the growth
exponent $\beta$ is the rate of change of interface width (or
roughness) in the transient regime ($w(t)\sim t^{\beta}$), the 
roughness exponent $\alpha$ shows the saturation of the width for 
a system with fixed size $L$ in the steady state regime 
($w(L) \sim L^{\alpha}$) and $z$ is the dynamical exponent. 
This model involves the traditional random deposition (at a 
rate of one complete monolayer during one unit of time) and 
surface relaxation such that the adatoms are searching for the 
sites with the minimum local height. We have taken the relaxation
length to be the lattice constant and we have applied the 
usual periodic boundary conditions. Typical sizes (i.e., number 
of lattice sites) used in this numerical work are $100-900$, 
and the averaging procedure implies a number of at least 
$10^5$ independent runs. All the measurements correspond 
to the steady state regime where the interface roughness
has reached a time independent equilibrium value (i.e., $t_0 \gg L^z$
in Eq.~(\ref{def})). We also mention that the smallest value 
for the sampling time is $1$. We emphasize that our use of Family
model is just a matter of convenience in simulating the EW equation
\cite{rev2}; our results are simply an exact discrete stochastic
simulation of the EW equation.

\begin{figure}
\includegraphics[height=7cm,width=8.5cm]{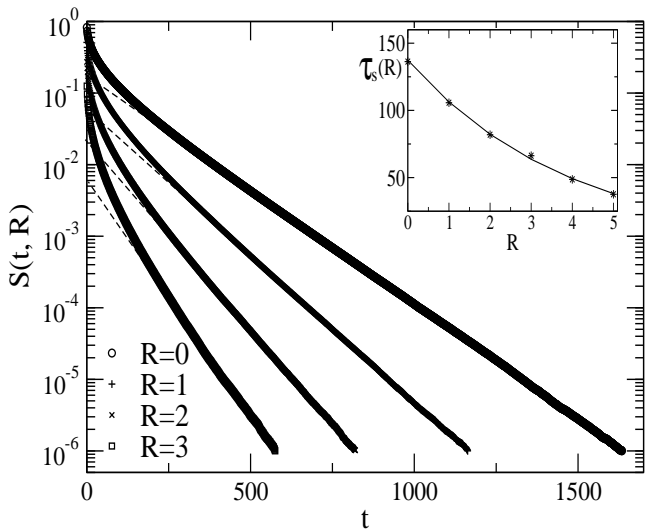} 
\caption{\label{fig1} The generalized survival probability, $S(t,R)$, 
for the discrete Family model. The dashed lines are fits of the long-time 
data to an exponential form. The system size is $L=100$, the sampling
time is $\delta t=1.0$ and the reference level $R$ takes four
different values: $0$, $1$, $2$ and $3$ (from top to
bottom). The inset shows the dependence of the generalized survival 
time scale $\tau_s(R)$ on the reference level value (up to $R=5$). 
The continuous curve represents a fit to an exponential decay of 
the generalized time scale vs. $R$. }
\end{figure}

\begin{figure}
\includegraphics[height=18cm,width=8.5cm]{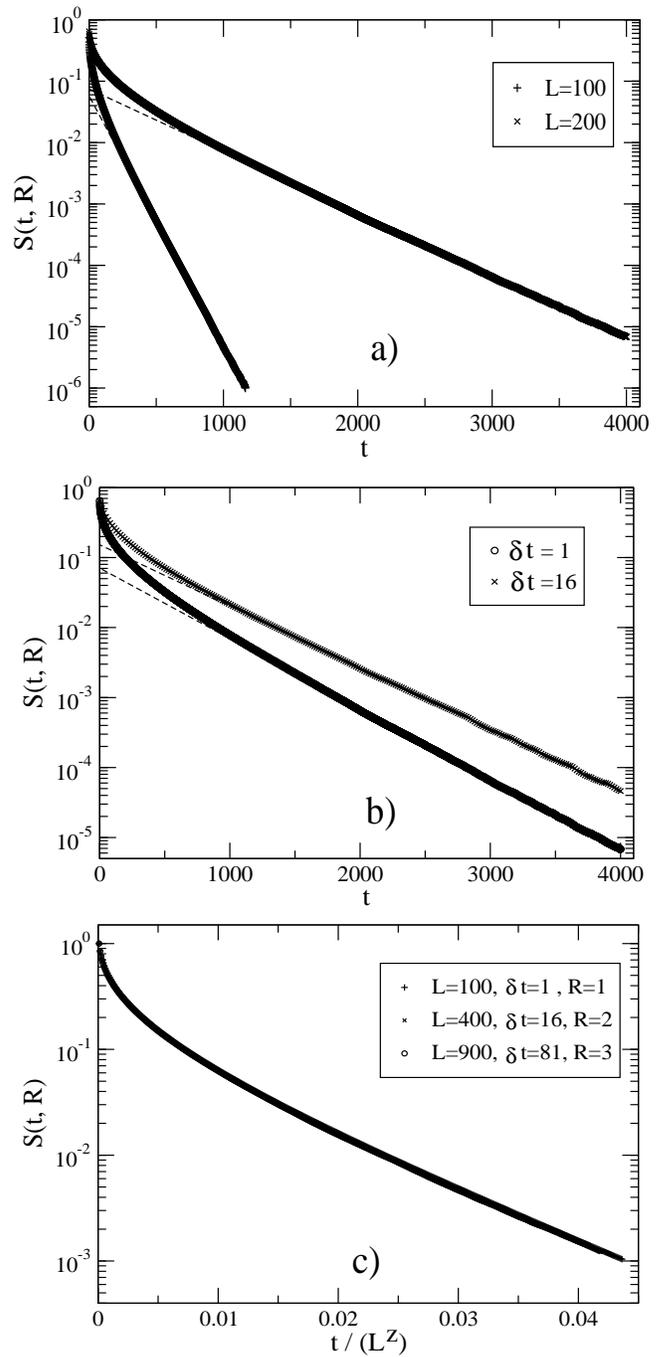} 
\caption{\label{fig2}
The generalized survival probability, $S(t,R)$, for the Family model. 
The dashed lines are fits of the long-time data to an exponential form.
Panel (a): $L=100$ (lower curve) and $L=200$ (upper curve), using 
fixed sampling time $\delta t=1$ and reference level $R=1$. 
Panel (b): $\delta t=1$ (lower plot) and $\delta t=16$ (upper plot) 
with fixed system size $L=200$ and reference level position $R=1$. 
Panel (c): Scaling of $S(t,R)$ using three different system sizes 
with sampling times and reference levels varied such that 
$\delta t/L^z$ and $R/L^{\alpha}$ are kept constant (i.e. 
$\delta t/L^z=1/10000$ and $R/L^{\alpha}=1/\sqrt{100}$). 
A perfect collapse of the curves with L=100, 400 and 900, 
respectively, occurs when using $z=2.03$.}
\end{figure}

Our results for the generalized survival probability and the
associated time scale are presented in Figs.~\ref{fig1} and
\ref{fig2}. $S(t,R)$ is simply computed as the fraction of sites
which, starting above (below) the level $R$ ($-R$) at time $t_0$, have
not crossed the reference level up to a later time $t_0+t$. In
Fig.~\ref{fig1} we show that, as expected, the generalized survival
with respect to an arbitrary reference level $R$ follows an
exponential decay at large times. The only varying parameter in
Fig.~\ref{fig1} is the reference level $R$. We have considered six
values for $R$, $R=0,1,...,5$ (only the first four curves are
displayed due to the limitations imposed by the quality of the
statistics, since as $R$ increases it is less probable to have a
reasonable number of lattice sites with height variables above (below)
$R$ ($-R$)). The dashed lines are fits of the long-time data to an
exponential form, $S(t,R) \propto \exp(-t/\tau_s(R))$. The upper
curve has $R=0$ and corresponds to the usual survival probability
previously studied in Ref.~\cite{BR}. However, all the other curves
are new and they prove that the generalized survival probability
decays exponentially in the long-time limit, with an associated
time scale, $\tau_s(R)$, which decreases with the reference level
value. As shown in the inset of Fig.~\ref{fig1}, the dependence of
$\tau_s(R)$ on $R$ is exponential, but clearly more work is needed in
order to understand this trend.  

In Fig.~\ref{fig2} we have used several lattice sizes, sampling times
and reference levels in order to identify the scaling behavior of
$S(t,R)$. In panel (a) we show the generalized survival with respect
to level $R=1$, measured using $\delta t=1$, for two system sizes:
$L=100$ and $L=200$. We observe that the underlying
survival time scale increases rapidly with $L$. In fact, $\tau_s(R)$
for a fixed $R$ is expected to grow proportionally to $L^z$
\cite{BR}. However, we obtain that $\tau_s(R=1)\simeq 103$ for $L=100$,
and $\tau_s(R=1)\simeq 429$ for $L=200$, so the measured generalized
survival time exhibits a small deviation from the expected value of 
$103 \times 4=412$. We find that this small effect is due to the dependence 
of the generalized survival on sampling time $\delta t$. This is 
clearly seen in panel (b). It turns out that a system with a fixed 
size ($L=200$) is characterized by different values of $\tau_s(R)$ 
if the sampling time of the measurement is adjusted. We observe that 
$\tau_s(R)$ increases weakly as the sampling time is increased. 
One might argue that this effect is very small and could be
neglected, but we have found that the effect of the sampling time on
the measured generalized survival probability has to be taken into
account in order to find the complete scaling function of $S(t,R)$. 
In addition, this effect is even stronger for systems with slower 
dynamics (i.e. larger $z$) \cite{BR}. Interestingly enough, we note 
that fixing the reference height level in the generalized survival 
probability problem introduces an additional length scale, that 
is related to the steady state value of the interface width, i.e. 
$L^{\alpha}$. Indeed, in panel (c) we look at three different 
systems with $L=200$, $400$ and $900$, respectively, and the 
generalized survival curves are calculated for $R=1$, $2$ and $3$,
respectively, i.e the level $R$ is varied proportionally to
$L^{\alpha}$, with $\alpha=1/2$ as appropriate for the EW 
equation. In addition, the sampling time 
for each of these three cases is also varied, 
$\delta t \propto L^z$ ($z=2$), so we have considered
$\delta t=1$ for $L=100$, $\delta t=16$ for $L=400$, and $\delta t=81$ 
for $L=900$, respectively. A perfect collapse of the curves $S(t,R)$
vs. $t/L^z$ occurs when using $z=2.03$, which agrees with the expected
value $z=2$, characteristic for the EW dynamics.

This numerical analysis allows us to conclude that the scaling form of
the generalized survival probability is
\begin{equation}
S(t,L,R,\delta t) = f(t/L^z,R/L^{\alpha},\delta t/L^z),
\label{scaling}
\end{equation}
where the function $f(x,y,z)$ decays exponentially for large values 
of $x$. The rate of this decay decreases rather rapidly as $y$ is 
decreased and increases rather slowly as $z$ is decreased. Note that 
for $y=0$ we recover the scaling form of the usual survival
probability with $R=0$ \cite{BR}.

To conclude, we have shown that the generalized 
survival probability of equilibrium step fluctuations
on vicinal surfaces with Edwards--Wilkinson dynamics decays 
exponentially at long times. We have investigated the
associated generalized survival time scale that depends on the 
system size $L$, sampling time $\delta t$, and the choice 
of the reference level $R$. In particular, the depedence of
$\tau_s(R)$ on $R$, which based on our preliminary investigations
seems to have an exponential trend, should be useful 
in understanding the stability of thermally 
fluctuating interfaces. We have also shown that the generalized 
survival probability exhibits simple scaling as a function of $L$, 
$\delta t$, and $R$. Our numerical results on $S(t,R)$ can be 
easily extended to fluctuating interfaces characterized by 
different dynamical evolutions (such as low--temperature 
step edge diffusion limited kinetics) belonging to different 
universality classes. Our goal here, using the example of the step
fluctuations process characterized by the EW universality class, is to
establish the generalized survival probability as an important
statistical concept in studying thermally fluctuating interfaces.

Finally, we mention that the generalized survival probability could be 
experimentally measured using dynamical STM step fluctuations 
data, opening the possibility for a direct approach to the crucial 
issue of interfacial stability. Our theoretical considerations for 
$S(t,R)$ should also be useful in understanding the dynamical evolution
of other physical processes \cite{exp} where a first-passage 
statistics has proven to be an useful concept.

The authors gratefully acknowledge discussions with E.D. Williams 
and C. Dasgupta. This work is partially supported by NSF-DMR-MRSEC 
and US-ONR.


\end{document}